\documentclass{jps-cp}
\usepackage{txfonts} %Please comment out this line unless the txfonts package is availabe in your LaTeX system.

\usepackage{bm}
\newcommand{\bea}{\begin{eqnarray}}
\newcommand{\eea}{\end{eqnarray}}
\newcommand{\be}{\begin{equation}}
\newcommand{\ee}{\end{equation}}

\def\nostrocostruttino#1\over#2{\mathrel{\mathop{\kern 0pt \rlap {\hbox{$#1$}}} \hbox{\kern-.125em $#2$}}}
\def\sumint{\nostrocostruttino \sum \over {\displaystyle\int}}

\title{General Helicity Formalism for Two-hadron Production in $e^+e^-$ Collisions and the $\Lambda$ Polarizing Fragmentation Function}

\author{Umberto \textsc{D'Alesio}$^{1,2,*}$, Francesco \textsc{Murgia}$^{2}$ and Marco \textsc{Zaccheddu}$^{1,2}$}

\inst{$^{1}$Dipartimento di Fisica, Universit\`a di Cagliari, Cittadella Univ., I-09042 Monserrato (CA), Italy \\
$^{2}$INFN, Sezione di Cagliari, Cittadella Univ., I-09042 Monserrato (CA), Italy\\
$^{*}$ Speaker.}

\email{umberto.dalesio@ca.infn.it}

\recdate{February 13, 2022}

\abst{We present the complete structure of the azimuthal dependences and polarization observables for two-hadron production in $e^+e^-$ annihilation processes within a transverse momentum dependent (TMD) approach adopting the helicity formalism. The leading-twist TMD fragmentation functions (TMD-FFs) for spin-1/2 hadrons are fully accounted for. The role of the polarizing FF, together with its extraction from Belle data for the transverse polarization of $\Lambda$'s, are discussed as well. Finally, predictions for SIDIS processes at EIC are presented.}

\kword{TMD fragmentation functions, transverse polarization, polarizing FF}

\begin{document}
\maketitle

\section{Introduction}
Hadron production in $e^+e^-$ annihilation processes represents the cleanest way to access the parton-to-hadron fragmentation mechanism.
The leading order (LO) expressions for all leading-twist (LT) azimuthal dependences and polarization observables are derived, within a TMD helicity formalism, for $e^+e^-\to h_1 h_2 \, X$ processes, together with the TMD-FFs for spin-1/2 hadrons~\cite{DAlesio:2021dcx}.
We then present a fit of Belle data~\cite{Belle:2018ttu} for the transverse $\Lambda$ polarization, leading to the extraction of the polarizing FF (pFF)~\cite{DAlesio:2020wjq}: the distribution of a transversely polarized spin-1/2 hadron in the fragmentation of an unpolarized quark. Finally, we show some estimates for SIDIS processes.

\section{Formalism}
We consider the process $e^+e^-\to h_1 h_2\,X$, where $h_{1,2}$ are spin-1/2 hadrons produced almost back-to-back. In the \emph{hadron-frame} configuration, the $\hat z_L$-axis is in the opposite direction w.r.t.~the hadron $h_2$ and the $\widehat{x_Lz_L}$ lepton plane is determined by the lepton and the $h_2$ directions (with the $e^+e^-$ axis at angle $\theta$). The production plane is fixed by $\hat z_L$ and the direction of the hadron $h_1$, with transverse momentum $\bm{P}_{1T}$, at an angle $\phi_1$ w.r.t.~the lepton plane.
The master formula at LO is given by
\bea
&&\rho^{h_1,S_1}_{\lambda_{h_1},\lambda^{'}_{h_1}} \rho^{h_2, S_2}_{\lambda_{h_2},\lambda^{'}_{h_2}} %%%%
\frac{d\sigma^{e^+e^-\rightarrow h_1 h_2 X}}{d\!\cos\!\, \theta\, dz_1\, d^2\!\bm{p}_{\perp 1} dz_2 \, d^2\!\bm{p}_{\perp 2}}\nonumber\\
&&= \sum_{q} \sum_{\{ \lambda \}} \frac{1}{32 \pi s}\frac{1}{4} %%%%%%
 \hat{M}_{\lambda_{q}\lambda_{\bar q},\lambda_{+}\lambda_{-}}\hat{M}^{\ast}_{\lambda^{'}_{q}\lambda^{'}_{\bar q},\lambda_{+}\lambda_{-}}  \hat{D}^{\lambda_{h_1},\lambda^{'}_{h_1}}_{\lambda_{q},\lambda^{'}_{q}}(z_1,\bm{p}_{\perp 1})  \hat{D}^{\lambda_{h_2},\lambda^{'}_{h_2}}_{\lambda_{\bar q},\lambda^{'}_{\bar q}}(z_2,\bm{p}_{\perp 2})\,,
\label{ee_hh}
\eea
where $z_i$ are the hadron light-cone momentum fractions and $\bm{p}_{\perp i}$ are the hadron transverse momenta w.r.t.~the parent parton directions of motion; $\rho^{h, S_h}_{\lambda_{h},\lambda^{'}_{h}}$ is the helicity density matrix of the hadron $h$ with spin $S_h$;  $\hat{M}_{\lambda_{q}\lambda_{\bar q},\lambda_{+}\lambda_{-}} $'s  are the helicity scattering amplitudes for the process $e^+(\lambda_+) + e^-(\lambda_-) \rightarrow q(\lambda_q) + \bar q(\lambda_{\bar q})$; $\hat{D}^{\lambda_{h},\lambda^{'}_{h}}_{\lambda_{q},\lambda^{'}_{q}}(z,\bm{p}_{\perp}) $ is the product of the helicity fragmentation amplitudes for the $q \rightarrow h +X$ process, defined as
\bea
\hat{D}^{\lambda_{h},\lambda^{'}_{h}}_{\lambda_{c},\lambda^{'}_{c}} (z,\bm{p}_{\perp}) &= & \sumint_{{X},\lambda_{X}}
\hat{\mathcal{D}}_{\lambda_{h},\lambda_{X};\lambda_c }(z,\bm{p}_\perp) \hat{\mathcal{D}}^{\ast}_{\lambda^{'}_{h},\lambda_{X};\lambda^{'}_c }(z,\bm{p}_\perp)\,,
\label{frag_mod}
\eea
where $\sumint_{{X},\lambda_{X}}$ stands for a spin sum and phase space integration over all undetected particles. All other details can be found in Ref.~\cite{DAlesio:2021dcx}.

The expression in Eq.~(\ref{ee_hh}) has then to be integrated over the unobserved variables as
\bea
\label{desahdfr2}
\frac{d\sigma^{e^+e^-\to h_1(S_1) h_2(S_2)\, X}}{d\!\cos\theta\, dz_1\,dz_2\,d^2\!\bm{P}_{1T}} %\nonumber\\
= \int\! d^2\!\bm{p}_{\perp 1} \,d^2\!\bm{p}_{\perp 2}\, \delta^{(2)}(\bm{p}_{\perp 1}- \bm{P}_{1T} + \bm{p}_{\perp 2}\,z_{p_1}/z_{p_2})\,
\frac{d\sigma^{e^+e^-\to h_1(S_1) h_2(S_2)\, X}}{d\!\cos\!\, \theta\, dz_1\, d^2\!\bm{p}_{\perp 1} dz_2 \, d^2\!\bm{p}_{\perp 2}}\,,
\eea
where $z_{p_{i}} = \frac{2 |\bm{P}_{h_{i}}|}{\sqrt s}$ are the momentum fractions ($z_{h_{i}} =\frac{2 E_{h_{i}}}{\sqrt s}$ are the usual energy fractions).

\subsection{Quark TMD-FFs for spin-1/2 hadrons}

We start defining the TMD-FF for a polarized quark, with spin $s_q$, fragmenting into an unpolarized hadron: $\hat{D}_{h/q,s_q}(z, \bm{p}_\perp)$. Then, we combine the parton and hadron helicity density matrices and the generalized FFs, Eq.~(\ref{frag_mod}),  as~\cite{DAlesio:2021dcx}
\begin{equation}%%%%quark
\rho^{h,S_h}_{\lambda_{h},\lambda^{'}_{h}} \hat{D}_{h/q,s_q}(z,\bm{p}_{\perp}) = \sum_{\lambda_{q},\lambda^{'}_{q}} \rho^{q}_{\lambda_{q},\lambda^{'}_{q}} \hat{D}^{\lambda_{h},\lambda^{'}_{h}}_{\lambda_{q},\lambda^{'}_{q}}(z,\bm{p}_{\perp})\,.
\label{hadron2quark}
\end{equation}
Exploiting the left-hand side in Eq.~(\ref{hadron2quark}), in terms of the polarization components, we can define
%\begin{equation}
% P^{h}_J \hat{D}_{h/q,s_q} = \hat{D}^{h/q}_{S_J/s_q}  -  \hat{D}^{h/q}_{-S_J/s_q} \equiv \Delta \hat{D}^{h/q}_{S_J/s_q}\,,
%\label{defP}
%\end{equation}
%
\bea
(P^h_J \, \hat D_{h/q,s_T}) &=&  \hat D_{S_J/s_T}^{h/q} - \hat D_{-S_J/s_T}^{h/q} \equiv
\Delta \hat D_{S_J/s_T}^{h/q}(z, \bm{p}_{\perp}) \label{DxY}\\
(P^h_J \, \hat D_{h/q,s_z}) &=&  \hat D_{S_J/s_z}^{h/q} -  \hat D_{-S_J/s_z}^{h/q} \; \equiv\Delta \hat D_{S_J/s_z}^{h/q}(z, \bm{p}_{\perp})  \label{DxZ}\\
\hat D_{h/q,s_T} &=&  D_{h/q}(z, p_{\perp}) + \frac{1}{2}\, \Delta \hat D_{h/s_T}(z, \bm{p}_{\perp})\,,\label{Dunp}
\label{main-tableq}
\eea
where $J=X,Y,Z$ ($P^{h}_J$) are the hadron helicity axes (polarizations), and $s_T$ ($s_z$) stands for the quark transverse (longitudinal) spin component in its helicity frame. These correspond to 8 TMD-FFs for a spin-1/2 hadron, with a clear partonic interpretation.

Taking into account the parity properties of $\hat{D}^{\lambda_{h},\lambda^{'}_{h}}_{\lambda_{q},\lambda^{'}_{q}}$, only 8 real independent quantities survive, directly related to the above 8 TMD-FFs~\cite{DAlesio:2021dcx}. These results can be recast as:
\bea
\hat{D}_{h/q}(z,\bm{p}_{\perp}) & = & D_{h/q}= (D^{++}_{++} +D^{++}_{--})\\
\Delta \hat{D}_{h/q,s_T}(z,\bm{p}_{\perp}) &  = & \Delta^{N}\! D_{h/q^{\uparrow}}\sin{(\phi_{s_q} - \phi_{h})} =  4 {\rm Im} D^{++}_{+-}\sin{(\phi_{s_q} - \phi_{h})}   \quad [{\rm Collins \ FF}]\\
\Delta \hat{D}^{h/q}_{S_Z/s_L}(z,\bm{p}_{\perp}) & = &\Delta D^{h/q}_{S_Z/s_L} =   (D^{++}_{++}- D^{++}_{--}) \\
\Delta \hat{D}^{h/q}_{S_Z/s_T} (z,\bm{p}_{\perp})&= &\Delta D^{h/q}_{S_Z/s_T}\cos{(\phi_{s_q} - \phi_{h})}= 2{\rm Re}D^{++}_{+-}\cos{(\phi_{s_q} - \phi_{h})} \\
\Delta \hat{D}^{h/q}_{S_X/s_L} (z,\bm{p}_{\perp})& = &\Delta D^{h/q}_{S_X/s_L}  = 2 {\rm Re}D^{+-}_{++}  \\
\Delta \hat{D}^{h/q}_{S_X/s_T}(z,\bm{p}_{\perp}) & = & \Delta D^{h/q}_{S_X/s_T}\cos{(\phi_{s_q} - \phi_{h})}= (D^{+-}_{+-} +D^{+-}_{-+})\cos{(\phi_{s_q} - \phi_{h})}\\
\Delta \hat{D}^{h}_{S_Y/q}(z,\bm{p}_{\perp}) & = &\Delta D^{h}_{S_Y/q} = \Delta^{N}\! D_{h^{\uparrow}/q} =
-2 {\rm Im}D^{+-}_{++}  \quad\quad\quad\quad\quad\quad\quad [{\rm Polarizing \ FF}] \\
\Delta^{-}\hat{D}^{h/q}_{S_Y/s_T}(z,\bm{p}_{\perp}) &= &  \Delta^{-}D^{h/q}_{S_Y/s_T} \sin{(\phi_{s_q} - \phi_{h})}=  (D^{+-}_{+-}-D^{+-}_{-+})\sin{(\phi_{s_q} - \phi_{h})}\,.
%}
\label{tab_ff_quark}
\eea

\subsection{Convolutions}

By fixing the hadron spins and summing over the helicity indices in Eq.~(\ref{ee_hh}), one obtains all azimuthal dependences in terms of pairs of TMD-FFs. A tensorial analysis allows to factor out the measurable azimuthal dependences and to express all quantities in terms of convolutions (see Ref.~\cite{DAlesio:2021dcx}):
\bea
\mathcal{C}[w \Delta D^{h_1}\Delta D^{h_2}] & = &   \sum_q e^2_q\!  \int\! d^2\!\bm{p}_{\bot 1} d^2\!\bm{p}_{\bot 2}\,\delta^{(2)} \big(\bm{p}_{\bot 1}  - \bm{P}_{ 1 T}  + \bm{p}_{\bot 2}
z_{p_{1}}/z_{p_{2}}  \big)\nonumber\\
&&\times\,
 w(\bm{p}_{\bot 2},\bm{P}_{1T})\,\Delta  D_{h_1/q}(z_1,p_{\bot 1}) \,\Delta D_{h_2/\bar q }(z_2,p_{\bot 2})
 \label{conv}\,.
\eea
Here we report only few examples (see also Refs.~\cite{Boer:1997mf,Pitonyak:2013dsu}):

- \emph{Unpolarized cross section}\\
\begin{equation}
\frac{d\sigma^{e^+e^-\rightarrow h_1 h_2 X}}{d\! \cos \theta\, dz_1  dz_2 d^2\!\bm{P}_{1T}} =  \frac{3 \pi \alpha^2}{2 s} \bigg\{ \big( 1 + \cos^2\theta \big)  F_{UU}   +  \sin^2\theta \cos(2 \phi_1) F^{\cos(2\phi_1)}_{UU} \bigg \}\,,
\end{equation}
with
\bea
F_{UU} &=&  \sum_q e^2_q  \int  d^2\bm{p}_{\bot 2}\, D_{h_1/q}(z_1, p_{\bot 1}) D_{h_2/\bar{q}}(z_2, p_{\bot 2}) = \mathcal{C}[D_{h_1/q} D_{h_2/\bar{q}}]\label{FUU1}\\
F^{\cos(2\phi_1)}_{UU}&=& \mathcal{C}\bigg[\frac{1}{4} \bigg\{\frac{P_{1T}}{p_{\perp 1}}\,\hat{\bm{p}}_{\bot 2} \cdot \hat{\bm{P}}_{1T}
- \frac{z_{p_1}}{z_{p_2}} \frac{p_{\perp 2}}{p_{\perp 1}}\,\bigg[ 2 \big(\hat{\bm{p}}_{\bot 2} \cdot \hat{\bm{P}}_{1T} \big)^2  - 1 \bigg]\bigg\} \Delta^N\! D_{h_1/q^\uparrow} \Delta^N\! D_{h_2/\bar{q}^\uparrow}\bigg] \,,
\label{FUUcos}
\eea
where the latter expression gives access to the Collins FF.

- \emph{Single-transverse polarized cross section}
\bea
&&P^{h_1}_T\, \frac{d\sigma^{e^+e^-\rightarrow h_1 h_2 X}}{d\! \cos \theta\, dz_1  dz_2 d^2\!\bm{P}_{1T}}= \frac{3\pi \alpha^2 }{2s}  \bigg\{ \Big( 1 + \cos^2\theta \Big)  \sin(\phi_1 - \phi_{S_1}^L)\, F^{\sin(\phi_1 - \phi_{S_1}^L)}_{TU}\nonumber\\
&&\quad +\, \sin^2\theta \,\bigg(  \sin(\phi_1 + \phi_{S_1}^L)\, F^{ \sin(\phi_1 + \phi_{S_1}^L)}_{TU} +   \sin(3\phi_1 - \phi_{S_1}^L)\, F^{\sin(3\phi_1 - \phi_{S_1}^L) }_{TU} \bigg) \bigg\}\,,\label{PTlab}
\eea
where $\phi_{S_1}^L$ is the azimuthal angle of the spin of the hadron $h_1$ and
\bea
F^{\sin(\phi_1 - \phi_{S_1}^L)}_{TU} & =&
\mathcal{C}\bigg[ \bigg( \frac{z_{p_1}}{z_{p_2}} \frac{p_{\bot2}}{ p_{\bot1}} \,\hat{\bm{p}}_{\bot 2} \cdot \hat{\bm{P}}_{1T} - \frac{P_{1T}}{p_{\perp 1}} \bigg)\, \Delta^N\! D_{h_1^\uparrow/q} D_{h_2/\bar{q}} \bigg]
\label{FTU1}\\
2 F^{ \sin(\phi_1 + \phi_{S_1}^L)}_{TU} &=&
\mathcal{C}\bigg [ \big( \hat{\bm{p}}_{\bot 2} \cdot \hat{\bm{P}}_{1T} \big) \frac{1}{2}\big(\Delta D^{h_1/q}_{S_X/s_T} + \Delta^-\! D^{h_1/q}_{S_Y/s_T}\big)\, \Delta^{N}\! D_{h_2/\bar{q}^{\uparrow}} \bigg ]
\label{FTU2}\\
2F^{\sin(3\phi_1 - \phi_{S_1}^L) }_{TU} &=&
\mathcal{C} \Bigg[\bigg\{ \frac{z_{p_1}^2}{z_{p_2}^2} \frac{p_{\perp 2}^2}{p_{\perp 1}^2}\bigg[
4\big( \hat{\bm{p}}_{\bot 2} \cdot \hat{\bm{P}}_{1T} \big)^3  - 3\, \big( \hat{\bm{p}}_{\bot 2} \cdot \hat{\bm{P}}_{1T} \big)\bigg]  + \frac{P_{1T}^2}{p_{\perp 1}^2}\,
 \big( \hat{\bm{p}}_{\bot 2} \cdot \hat{\bm{P}}_{1T} \big)  \nonumber \\
&&\;- 2\, \frac{z_{p_1}}{z_{p_2}}\frac{p_{\perp 2} P_{1T}}{p_{\perp 1}^2}\bigg[ 2\big( \hat{\bm{p}}_{\bot 2} \cdot \hat{\bm{P}}_{1T} \big)^2 - 1 \bigg]  \bigg\}
\frac{1}{2} \bigg(\Delta D^{h_1/q}_{S_X/s_T} - \Delta^-\! D^{h_1/q}_{S_Y/s_T}\bigg)\, \Delta^{N}\! D_{h_2/\bar{q}^{\uparrow}}  \,.\label{FTU3}
\eea

The polarization orthogonal to the hadron plane, that is along
\be
\hat{\bm{n}} \equiv (\cos\phi_n, \sin\phi_n,0)= \frac{- \bm{P}_2  \times  \bm{P}_1}{|\bm{P}_2  \times  \bm{P}_1|} = -\sin\phi_1 \hat{\bm{x}}_L + \cos\phi_1 \hat{\bm{y}}_L\,,
\ee
can be obtained by identifying $\phi_{S_1}^L\equiv\phi_n=\phi_1 + \frac{\pi}{2}$ in Eq.~(\ref{PTlab}). By  integrating over $\bm{P}_{1T}$ we get
\be
P^{h_1}_n(z_1,z_2) = -\frac{F^{\sin(\phi_1 - \phi_{S_1}^L)}_{TU}}{F_{UU}} \,.
\label{Pn1}
\ee

\section{Phenomenology}

By adopting a Gaussian Ansatz for the TMD-FFs in Eqs.~(\ref{FUU1}), (\ref{FTU1}), Eq.~(\ref{Pn1}) becomes
\bea
P^{h_1}_n = \sqrt{\frac{e\pi}{2}}\frac{1}{M_{\rm pol}} \frac{\langle p_\perp^2\rangle_{\rm pol}^2}{\langle p_{\perp 1}^2\rangle}\,\frac{z_2}{\Big\{\Big[z_1\Big(1-\frac{m_{h_1}^2}{z_1^2s}\Big)\Big]^2\langle p_{\perp 2}^2\rangle +z_2^2 \langle p_\perp^2\rangle_{\rm pol}\Big\}^{1/2}} %\nonumber\\
%&\times &
\,\frac{\sum_{q} e^2_q\,\Delta D_{h_1^\uparrow/q}(z_1)D_{h_2/\bar q}(z_2)}{ \sum_{q}  e^2_q\,
 D_{h_1/q}(z_1)D_{h_2/\bar q}(z_2)}\,,
\label{PolTh}
\eea
where we have used the following parametrizations:
\bea
D_{h/q}(z,p_{\perp}) =  D_{h/q}(z)\, \frac{e^{-p_{\perp}^2/\langle p_{\perp}^2\rangle}}{\pi \langle p_{\perp}^2\rangle} \quad\quad\quad
\Delta^N\! D_{h^\uparrow\!/q}(z,p_{\perp}) =  \Delta D_{h^\uparrow\!/q}(z)  \frac{\sqrt{2e} \,p_{\perp}}{M_{\rm pol}} \frac{e^{-p_{\perp}^2/\langle p_{\perp}^2\rangle_{\rm pol}}}{\pi \langle p_{\perp}^2\rangle}\,,
\label{Gaus}
\eea
with $\langle p_\perp^2\rangle_{\rm pol} = \frac{M_{\rm pol}^2}{M_{\rm pol}^2 + \langle p_{\perp}^2\rangle} \,\langle p_{\perp}^2\rangle$.
Another important quantity is the first $p_\perp$-moment of the pFF:
\be
\Delta D_{h^\uparrow\!/q}^{(1)}(z)  =  \int\! d^2 \bm{p}_{\perp} \frac{p_{\perp}}{2 z m_h} \Delta^N\! D_{h^\uparrow\!/q}(z,p_{\perp})
%= D_{1T}^{\perp (1)}(z)
= \sqrt{\frac{e}{2}}\frac{1}{z m_h} \frac{1}{M_{\rm pol}}\frac{\langle p^2_{\perp} \rangle^{2}_{\rm pol}}{\langle p^2_{\perp} \rangle}  \Delta D_{h^\uparrow\!/q }(z)\,.
\label{1mom}
\ee

We also consider the inclusive $\Lambda$ production case (within a jet), adopting a simplified approach in terms of TMD-FFs. The transverse polarization (w.r.t.~the jet-$\Lambda$ plane) is given as
\bea
{\cal P}_T(z, p_{\perp}) &  = & \frac{\sum_{q} e^2_q\,
\Delta^N\! D_{\Lambda^\uparrow/q}(z,p_{\perp})}{ \sum_{q}  e^2_q\,
D_{\Lambda/q}(z,p_{\perp})}\,.
\label{PolTjet}
\eea

\subsection{Results}
\label{results}

We first perform a fit of the associated production data alone (see also Ref.~\cite{Callos:2020qtu}), including in a second phase also the inclusive data set~\cite{DAlesio:2020wjq}. The $z$-dependent part of the pFF is parameterized as follows ($q=u,d,s, \rm{sea}$)
\be
\Delta D_{\Lambda^\uparrow\!/q}(z) = N_q z^{a_q} (1-z)^{b_q} \frac{(a_q+b_q)^{(a_q+b_q)}}{a_q^{a_q}b_q^{b_q}} D_{\Lambda/q}(z)\,,
\label{Deltaz}
\ee
where $|N_q|\le 1$. The best parameter choice turns out to be:
$ N_u, \; N_d, \; N_s, \; N_{\rm sea}, \; a_s, \; b_u, \; b_{\rm sea}\,,
$
with all other $a$, $b$ parameters set to zero. Together with $\langle p^2_{\perp} \rangle_{\rm pol}$ (Eq.~(\ref{Gaus})) we have 8 free parameters.

\begin{figure}[!t]
\centering
\begin{minipage}[b]{0.48\textwidth}
\includegraphics[width=7.cm]{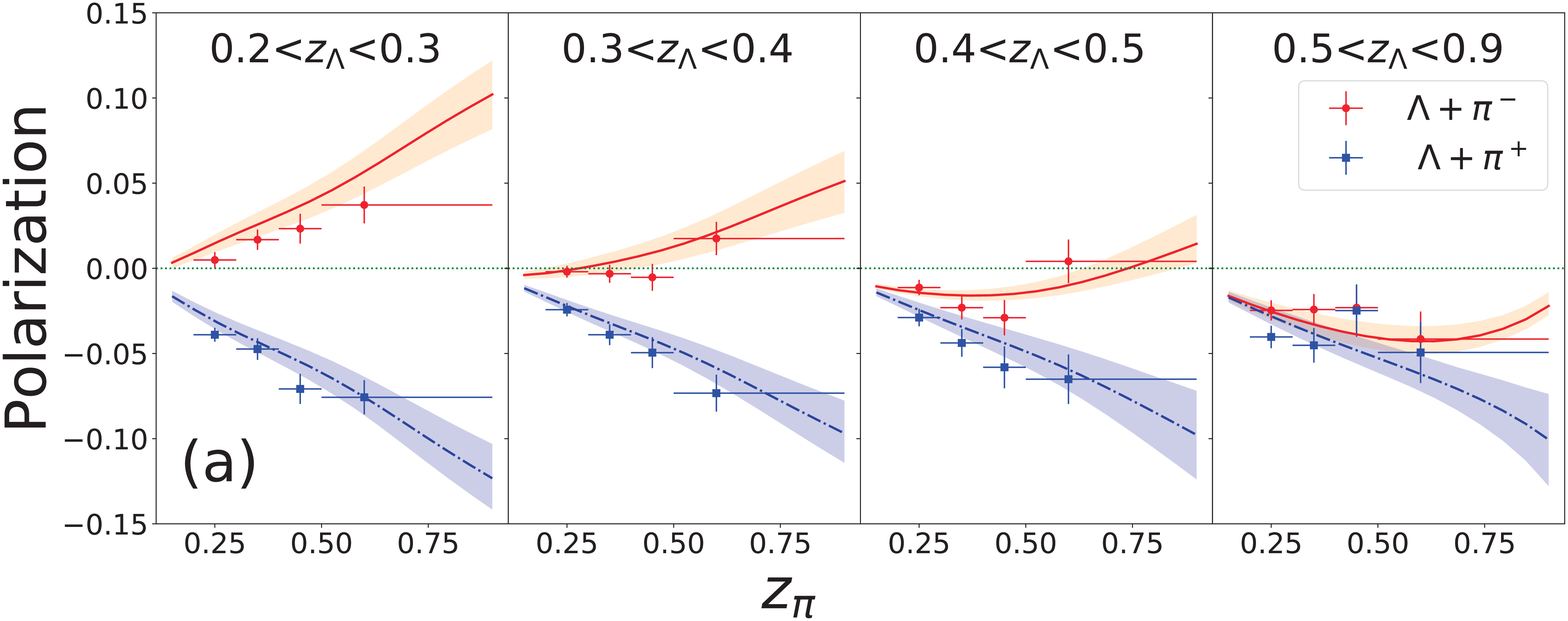}
\end{minipage}
\begin{minipage}[b]{0.48\textwidth}
\includegraphics[width=7.cm]{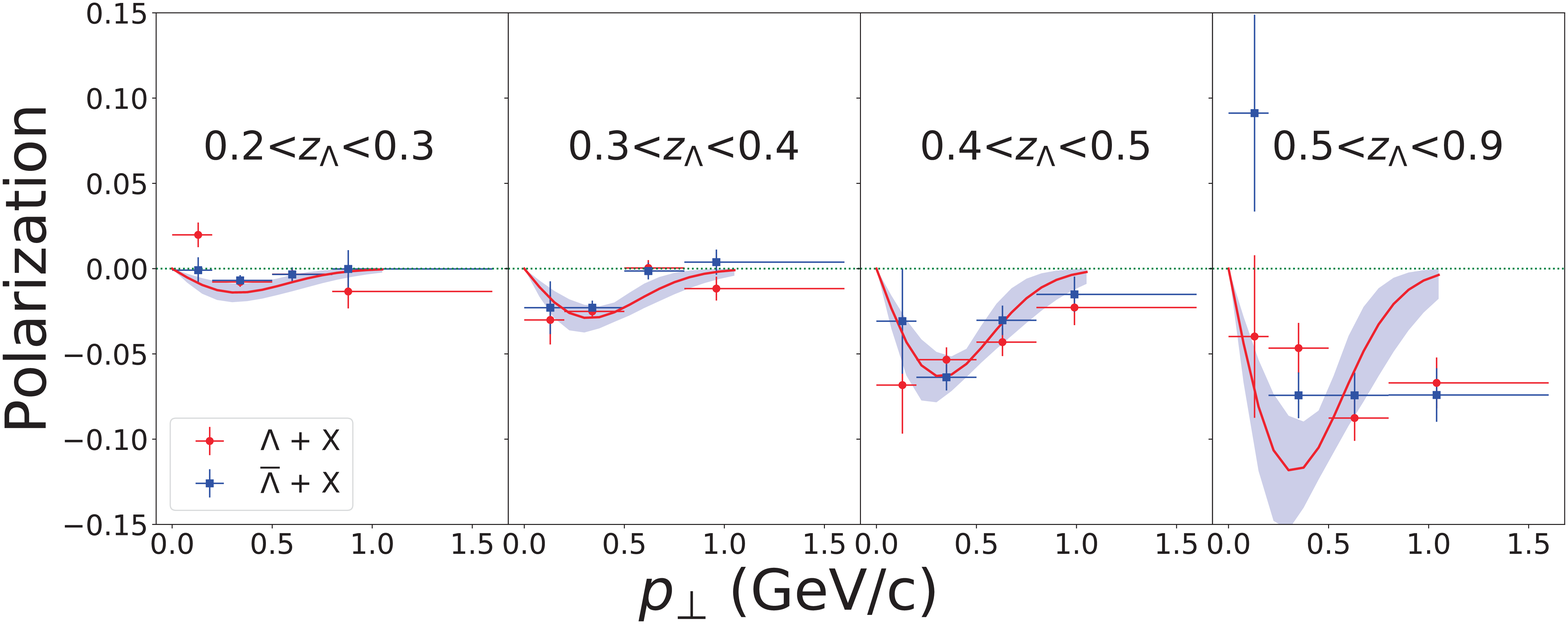}
\end{minipage}
\caption{Best-fit estimates, based on the Belle full-data set~\cite{Belle:2018ttu}, of the transverse polarization for $\Lambda$/$\bar\Lambda$ production in $e^+e^-\to \Lambda\, \pi \, X$ vs.~$z_{\pi}$ (left panel) and $e^+e^-\to \Lambda(\bar\Lambda) \, X$ vs.~$p_{\perp}$ (right panel), for different $z_\Lambda$ bins. }
\label{fig:Lhj}
\end{figure}
\begin{figure}[thb]
\centering
\includegraphics[trim =  150 0 300 60,clip,width=3.5cm]{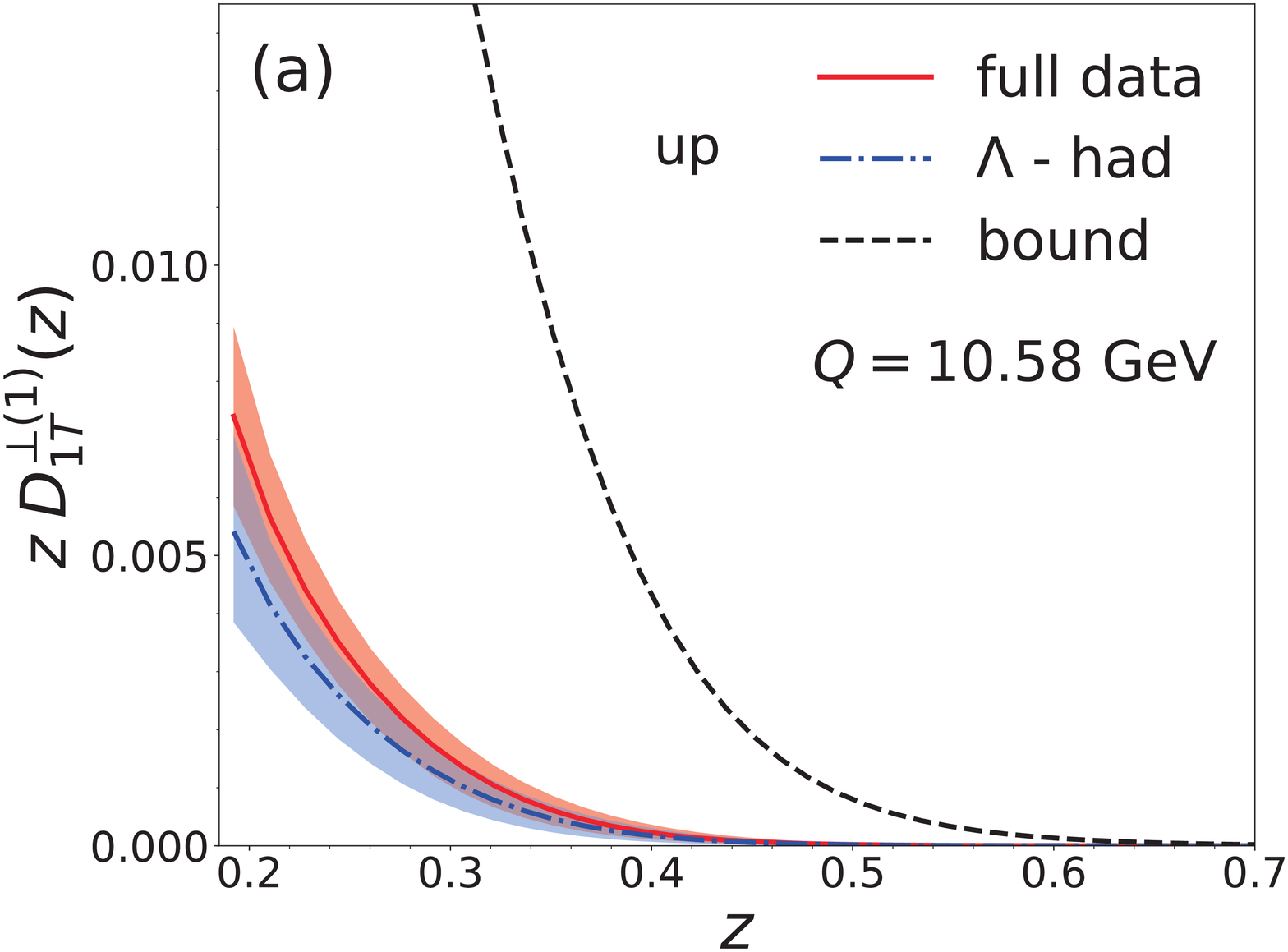}\;
\includegraphics[trim =  150 0 300 60,clip,width=3.5cm]{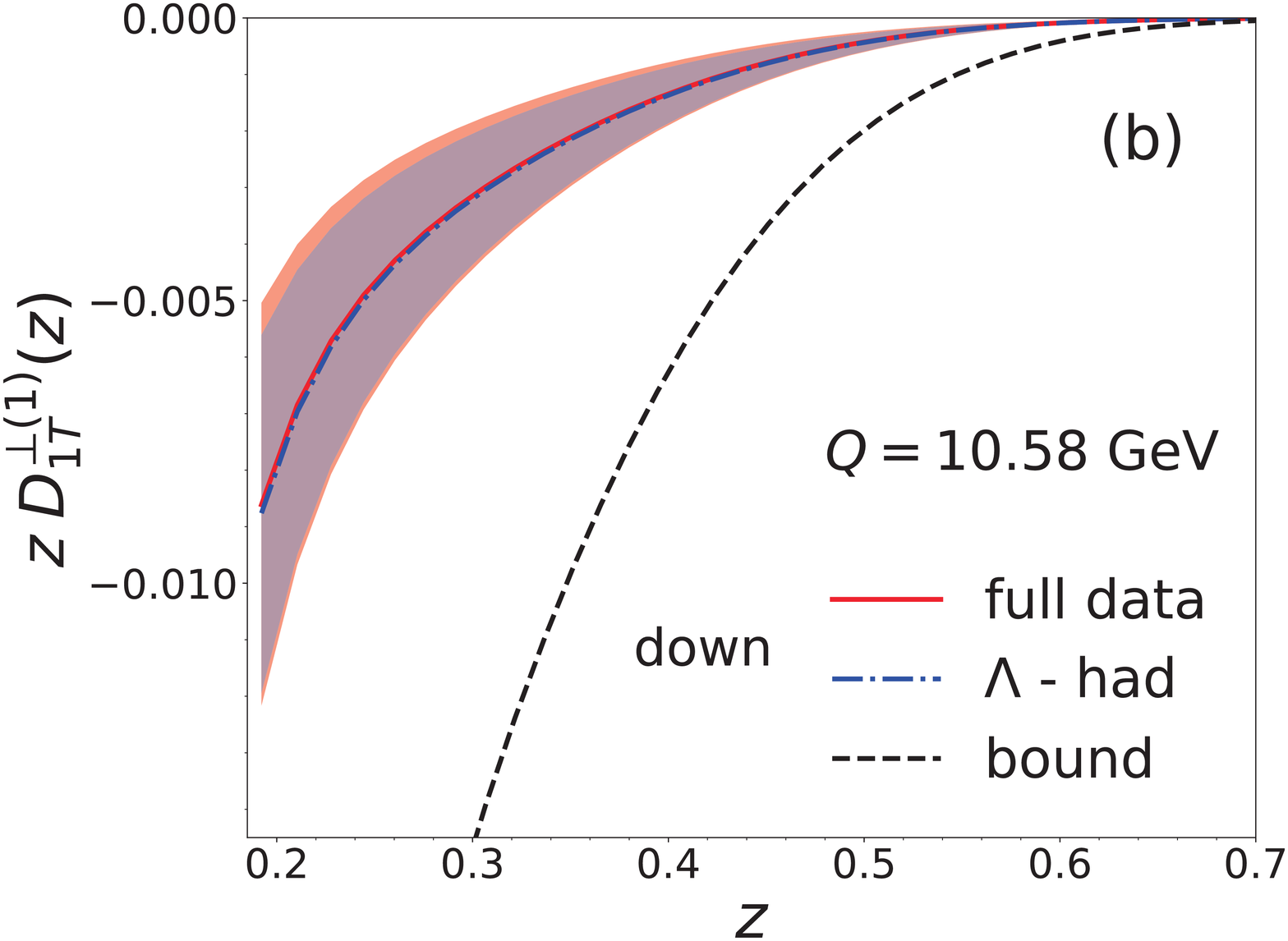}\;
\includegraphics[trim =  150 0 300 60,clip,width=3.5cm]{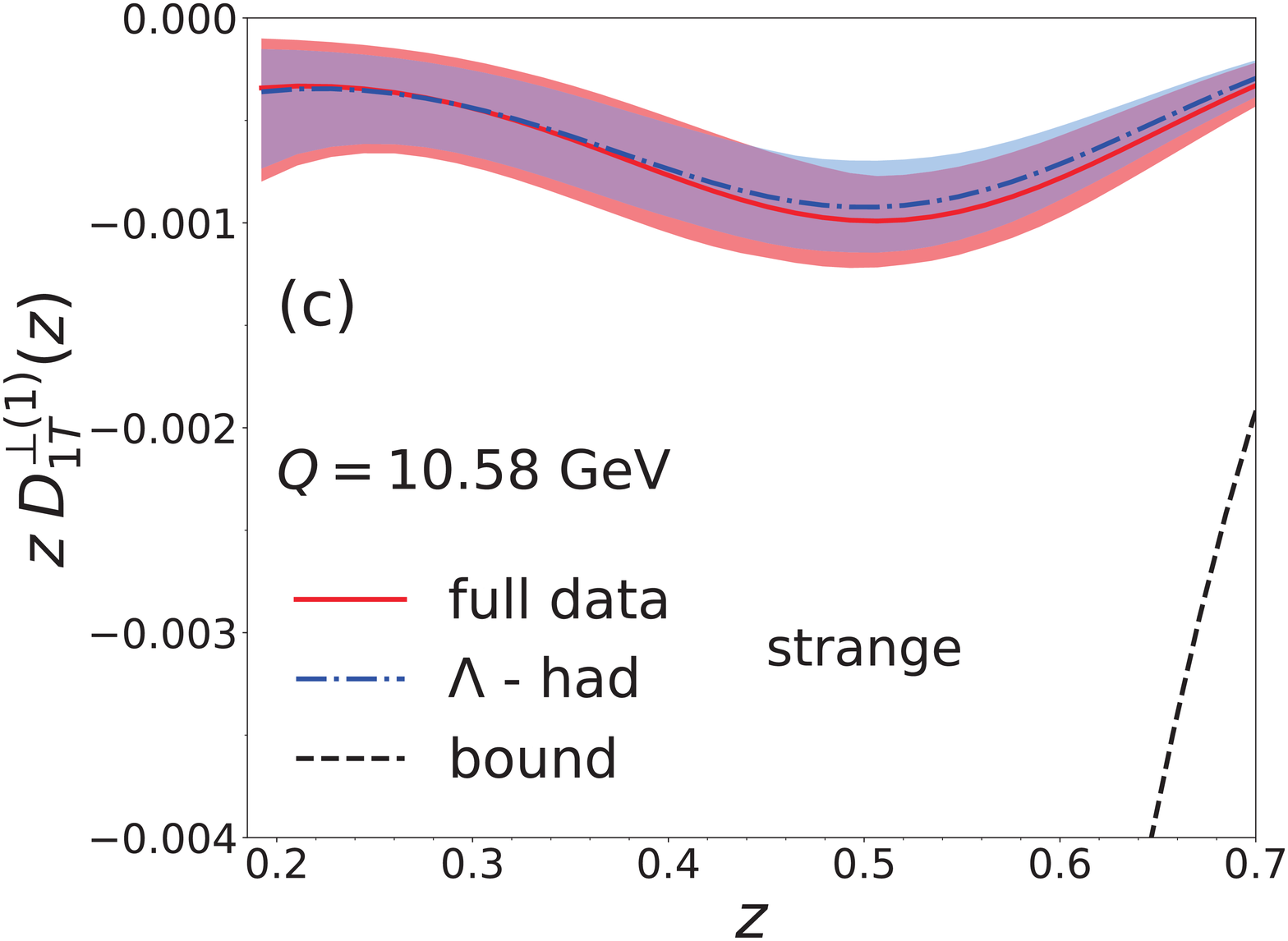}\;
\includegraphics[trim =  150 0 300 60,clip,width=3.5cm]{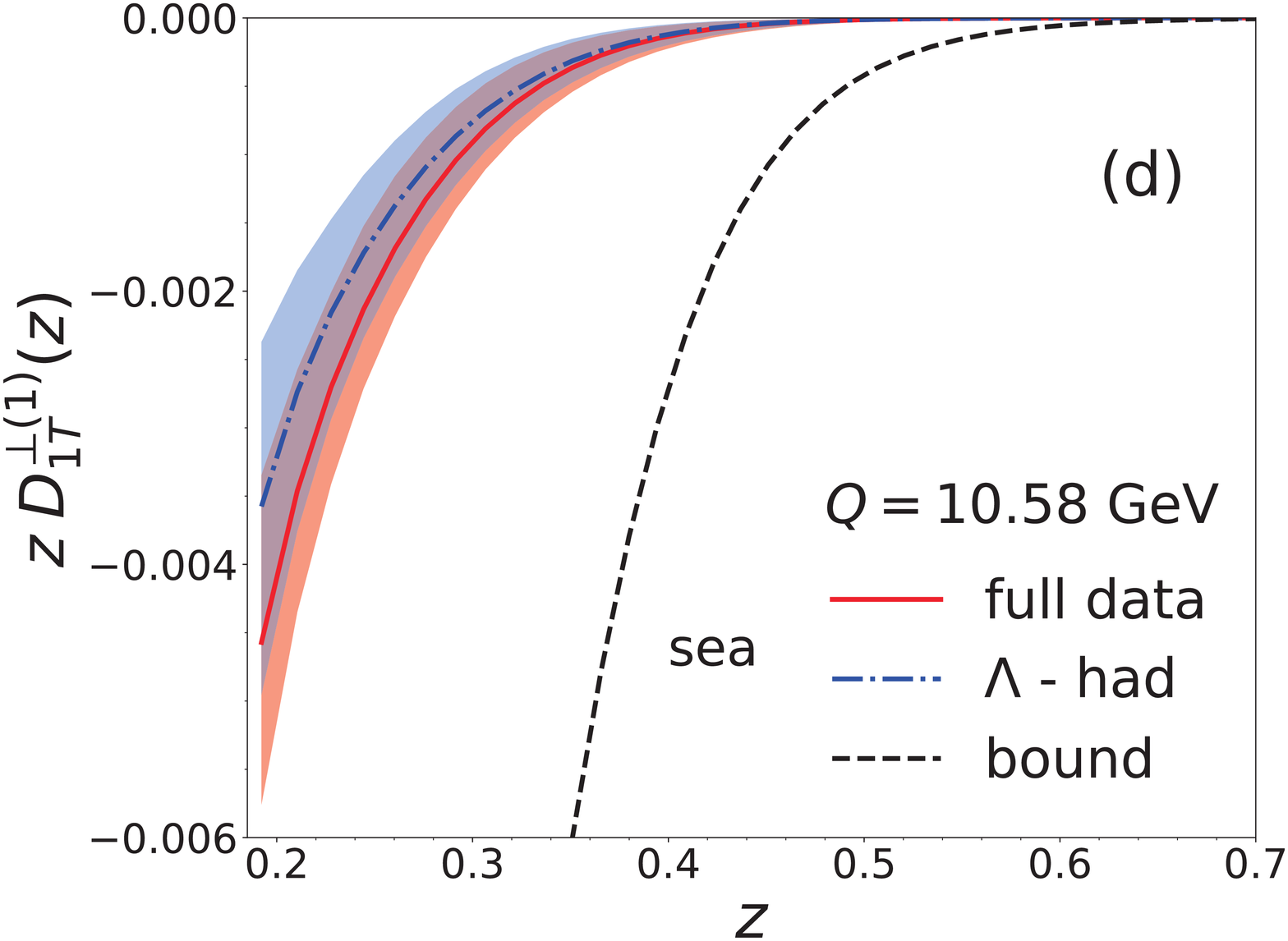}
\caption{First moments of the pFFs, Eq.~(\ref{1mom}), for the up (a), down (b), strange (c) and sea (d) quarks, as obtained from the full-data fit (red solid lines) and the $\Lambda$-hadron fit (blue dot-dashed lines).}
\label{fig:1stm}
\end{figure}

Some estimates (with their statistical uncertainty bands at $2\sigma$-level), compared against Belle data~\cite{Belle:2018ttu}, are shown in Fig.~\ref{fig:Lhj}. The associated production (full-) data fit leads to a $\chi^2_{\rm dof}$ = 1.26 (1.94), while the corresponding first moments, Eq.~(\ref{1mom}), are quite stable (Fig.~\ref{fig:1stm}).

\begin{figure}[!thb]
\centering
\includegraphics[width=6.cm]{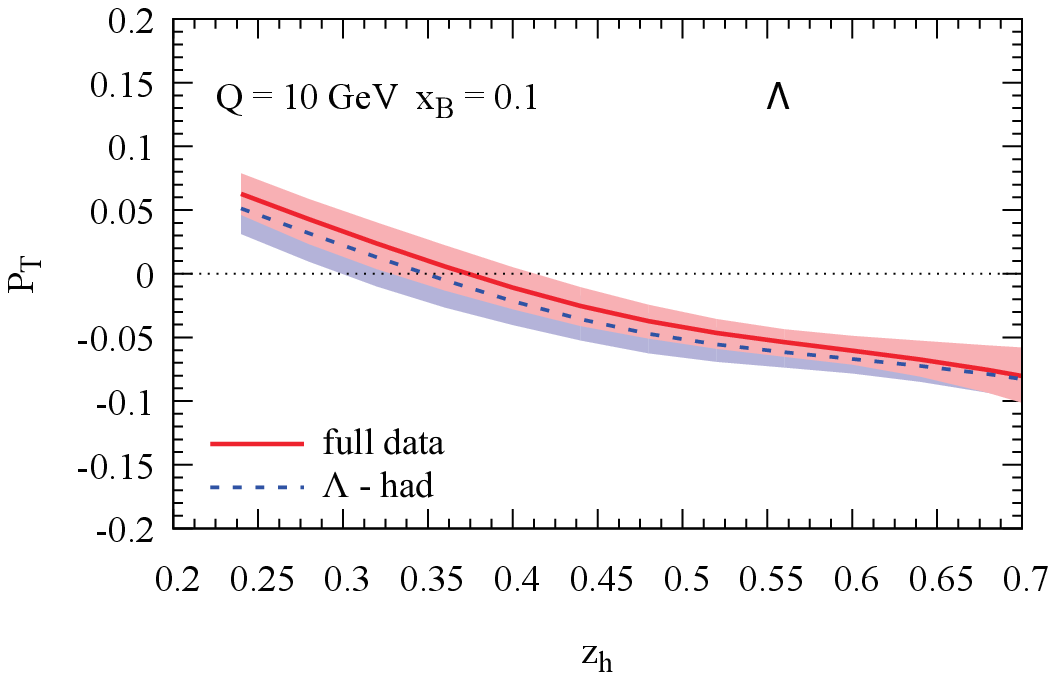}
\includegraphics[width=6.cm]{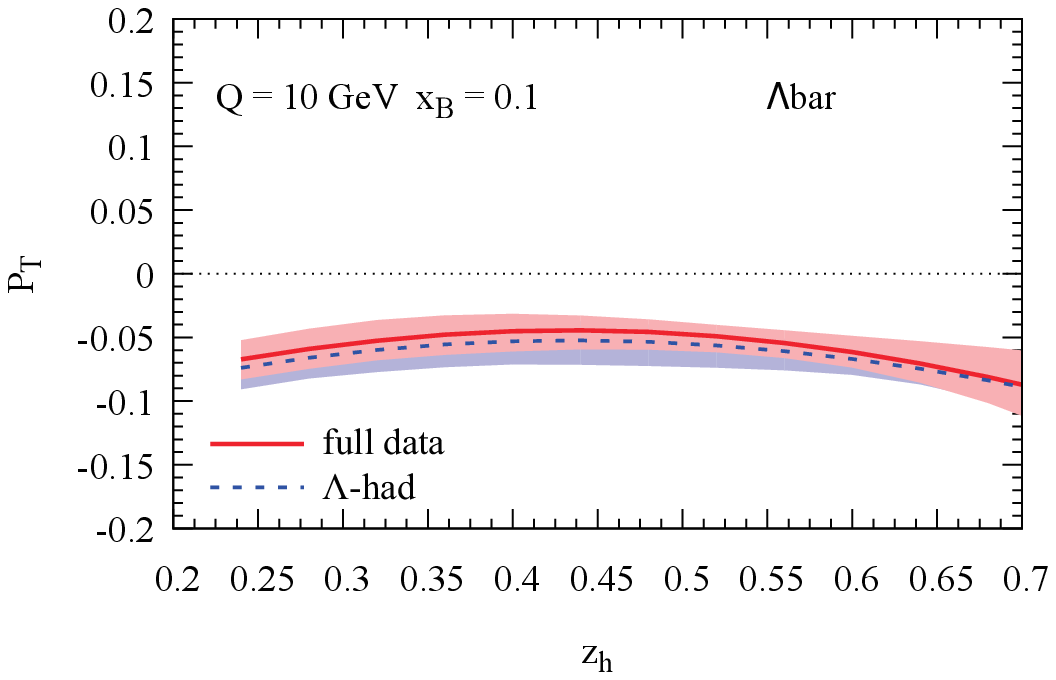}
\caption{Estimates for the transverse polarization of $\Lambda$/$\bar\Lambda$ in SIDIS for EIC kinematics (see legend).}
\label{fig:sidis}
\end{figure}

We now use the so extracted pFFs to give predictions for the same observable in SIDIS. In such a case the polarization is measured transversely w.r.t.~the plane containing the target and the $\Lambda$ particle. The final result, as a function of $x_{\rm B}$ and $z_h$ and adopting a Gaussian parametrization also for the unpolarized TMD parton distribution, reads:
\be
P_T(x_{\rm B},z_h) = \frac{\sqrt{2e\pi}}{2m_p}
\frac{\langle p_\perp^2\rangle^2_{\rm pol}}{\langle p_\perp^2\rangle}
\frac{1}{\sqrt {\langle p_\perp^2\rangle_{\rm pol} + z_h^2 \langle k_\perp^2\rangle }} \frac{\sum_q e_q^2 f_{q/p}(x_{\rm B})\Delta D_{\Lambda^\uparrow/q}(z_h)}{\sum_q e_q^2 f_{q/p}(x_{\rm B}) D_{\Lambda/q}(z_h)}\,.
\ee
The corresponding estimates for EIC kinematics are shown in Fig.~\ref{fig:sidis}. This would definitely allow for a test of the universality of the pFF as well as of its flavor dependence.

\section{Conclusions}
\label{concl}
The complete azimuthal structure in terms of TMD-FFs for the process $e^+e^-\to h_1 h_2\, X$ has been derived within the helicity formalism in full detail and the first extraction of the polarizing FF from Belle data has been presented. Estimates for the transverse $\Lambda$ polarization in SIDIS are given, emphasising their role in the study of the (universality) properties of the pFF.

%\bibliographystyle{unsrt}
%\bibliographystyle{h-physrev}
%\bibliography{references.bib}
\end{document}